\documentclass[5p]{elsarticle}

\usepackage[T1]{fontenc}
\usepackage{units}
\usepackage{setspace}
\usepackage{color}
\usepackage{amsmath}
\usepackage{psfrag}
\usepackage{subfigure}

\biboptions{sort&compress}

\usepackage{amssymb}

\journal{Physics Letters B}

\begin{document}

\begin{frontmatter}

\title{
\hfill {\small DESY 10-030, MPP-2010-27}\\
New ALPS Results on Hidden-Sector Lightweights}

\author[a]{Klaus Ehret}
\author[b]{Maik Frede}
\author[a]{Samvel Ghazaryan}
\author[b]{Matthias Hildebrandt}
\author[a]{Ernst-Axel Knabbe}
\author[b]{Dietmar Kracht}
\author[a]{Axel Lindner{\corref{cor1}}}
\author[a]{Jenny List}
\author[c]{Tobias Meier}
\author[a]{Niels Meyer}
\author[a]{Dieter Notz}
\author[a,d]{Javier Redondo}
\author[a]{Andreas Ringwald}
\author[e]{G\"unter~Wiedemann}
\author[c]{Benno Willke}

\cortext[cor1]{Corresponding author, e-mail: axel.lindner@desy.de}

\address[a]{Deutsches Elektronen-Synchrotron DESY, Notkestra\ss e 85, D-22607 Hamburg, Germany}
\address[b]{Laser Zentrum Hannover e.V., Hollerithallee 8, D-30419 Hannover, Germany}
\address[c]{Max-Planck-Institute for Gravitational Physics, Albert-Einstein-Institute, and Institut f\"ur Gravitationsphysik,
Leibniz Universit\"at, Hannover, Callinstra\ss e 38, D-30167 Hannover, Germany}
\address[d]{Present address: Max-Planck-Instit\"ut f\"ur Physik, F\"ohringer Ring 6, D-80805 M\"unchen, Germany}
\address[e]{Hamburger Sternwarte, Gojenbergsweg 112, D-21029 Hamburg, Germany}

\begin{abstract}
The ALPS collaboration runs a ``Light Shining through a Wall'' (LSW) experiment to search for photon oscillations into ``Weakly Interacting Sub-eV Particles'' (WISPs) often predicted by extensions of the Standard Model. 
The experiment is set up around a superconducting HERA dipole magnet at the site of DESY.
Due to several upgrades of the experiment we are able to place limits on the probability of photon-WISP-photon conversions of a few $\times 10^{-25}$. These limits result in 
today's most stringent laboratory constraints on the existence of low mass axion-like particles, hidden photons and minicharged particles.
\end{abstract}

\begin{keyword}
Experimental tests \sep photon regeneration \sep resonators and cavities \sep axions \sep other gauge bosons
\PACS 12.20.Fv \sep 14.80.-j \sep 14.80.Mz \sep 14.70.Pw \sep 42.60.Da
\end{keyword}
\end{frontmatter}

\newpage

\section{Introduction}

Despite of its tremendous phenomenological success, it is believed that the Standard Model of elementary particle physics
is not the ultimate theory of space, time and matter. In fact, there is a growing number of theoretical proposals for
extensions of the Standard Model (SM), trying to arrive at an even more unified description of particles and their
interactions. Very frequently, these extensions exhibit their truly unifying features only at very high energy scales,
far above the electroweak ($\sim$~TeV) scale currently probed by the LHC. However, at the same time,
these unified theories predict very often a rich low-energy phenomenology arising from very weakly interacting
sub-eV particles (WISPs) beyond the SM.

As the first and paradigmatic example we find the axion~\cite{Weinberg:1977ma}  
and other axion-like particles (ALPs),
the smallness of their mass being related to a shift symmetry of the field theory of the corresponding
axion field $\phi(x)$ under the transformation
$\phi(x)\to \phi(x) + {\rm const}$. Such a symmetry forbids explicit mass terms, $\propto m_\phi^2 \phi^2$,
in the Lagrangian, rendering the particles corresponding to the excitations of the field $\phi(x)$  massless.
Moreover, it leads to the fact that the couplings of axions and ALPs to
standard model particles can only occur via derivative couplings, $\propto \partial \phi/f_\phi$, leading to a strong
suppression of their interactions at energy scales below $f_\phi$, their decay constant,
promoting them to perfect WISP candidates~\cite{Kim:1979if}.

ALPs may have also non-zero, but still small masses, $m_\phi\sim \Lambda^2/f_\phi\ll f_\phi$, arising from terms in the low energy effective Lagrangian which break the shift symmetry explicitly at a scale $\Lambda\ll f_\phi$.
This occurs especially for the proper axion $a$
(sometimes called Peccei-Quinn axion or QCD axion) whose shift symmetry is broken by a QCD
colour anomaly~\cite{Peccei:1977hh}. 
The latter has the virtue of solving the ``strong CP problem''
by promoting the CP violating parameter $\theta$ in QCD essentially to a field $a(x)/f_a$,
which dynamically relaxes to zero, thereby explaining the apparent smallness of strong CP violation.
Moreover, for $f_a\sim 10^{12}$~GeV, the axions would be perfect candidates for the observed -- yet unidentified -- dark matter of the universe~\cite{Abbott:1982af}.
Besides the axion, unifying theories beyond the SM, in particular string theory, appear to predict a possibly large number
of ALPs~\cite{Svrcek:2006yi}.

\begin{figure*}[t] 
\centering
\includegraphics[width=16cm]{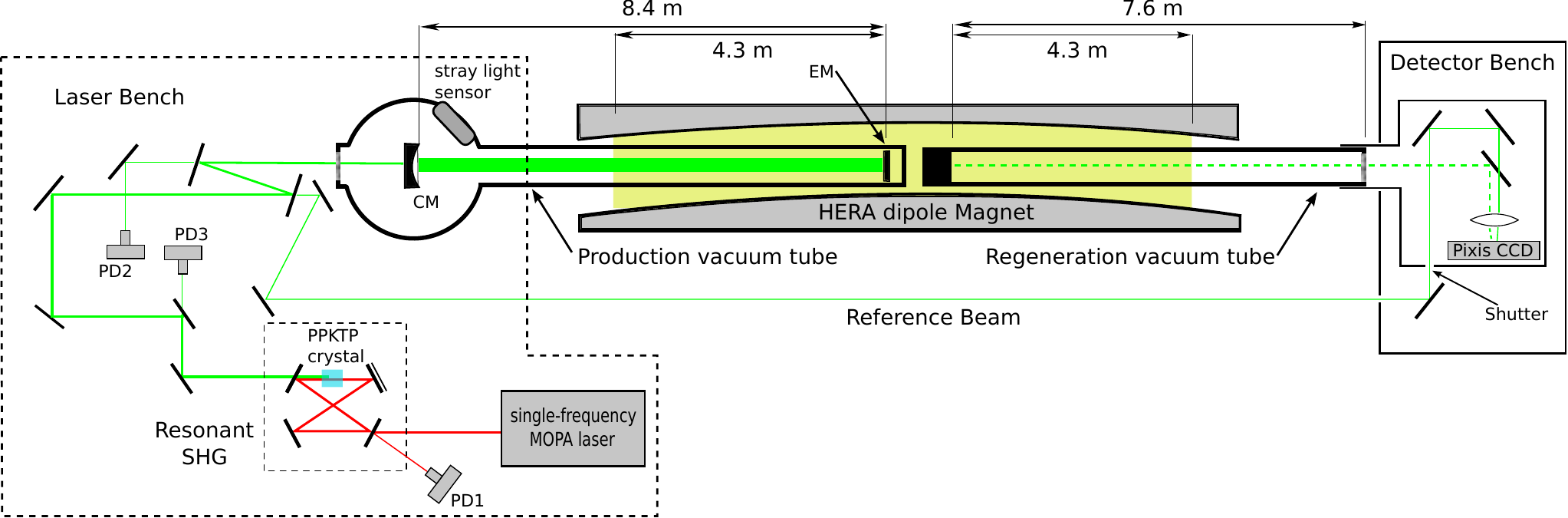}
\caption{Schematic view of the ALPS LSW experiment. See the text for a description. }
\label{fig:setup}
\end{figure*}

Phenomenologically most important is the ALP field coupling to two photons, which generically has the form
\begin{eqnarray}
\label{ALPcoupling_pseudoscalar}
{\cal L}_{\phi^{(-)}\gamma\gamma} &=&-\frac{g_-}{4} F_{\mu\nu}\widetilde F^{\mu\nu}\phi^{(-)} ,\\
\label{ALPcoupling_scalar}
{\cal L}_{\phi^{(+)}\gamma\gamma} &=&-\frac{g_+}{4} F_{\mu\nu} F^{\mu\nu}\phi^{(+)} ,
\end{eqnarray}
where $F_{\mu\nu}$ is the electromagnetic field strength, $\widetilde F_{\mu\nu}$ its dual, and $g_\pm\propto \alpha /f_\phi$,
with fine-structure constant $\alpha$.
The couplings \eqref{ALPcoupling_pseudoscalar} and \eqref{ALPcoupling_scalar} for pseudoscalar and scalar
ALPs, respectively, provide photon-ALP mixing in a background electromagnetic field~\cite{Raffelt:1987im},
which produces photon-ALP oscillations that can be searched for in a variety of contexts (for a recent review, see
Ref.~\cite{Jaeckel:2010ni}),
including dedicated laboratory searches, in particular in
light shining through a wall experiments~\cite{Okun:1982xi,Anselm:1986gz}. 
Interestingly, puzzling astronomical observations such as the recently reported anomalous transparency of the universe to gamma-rays, the alignment of distant quasar polarizations or the unexplained scatter in luminosity relations 
may find their explanation in terms of such a photon-ALP 
mixing~\cite{Csaki:2001yk}.
Similarly the fast cooling of white dwarfs may hint at an ALP coupling to electrons with a comparable strength~\cite{Isern:2008nt}.

The rich phenomenology of photon-ALP oscillations extends to other well-motivated WISPs, in particular to
so-called hidden photons~\cite{Okun:1982xi,Ahlers:2007rd}. In fact, light, extra hidden U(1) gauge bosons occur
very frequently in unified theories beyond the Standard Model~\cite{Goodsell:2010ie}. In general, these hidden photons  will
mix with the standard photon via a kinetic mixing term,
\begin{equation}
\label{kinmix}
{\cal L}_{\gamma'\gamma} = -\frac{1}{2}\chi F_{\mu\nu}B^{\mu\nu},
\end{equation}
where $B_{\mu\nu}$ is the field strength of the hidden photon, which can obtain a mass from either
a hidden Higgs or a Stueckelberg mechanism. The predicted values of the kinetic mixing, $\chi$, are widespread and depend strongly on
the particular extension of the SM.  Typical numbers in string embeddings of the SM lie in the broad range $10^{-16}\sim10^{-2}$~\cite{Dienes:1996zr,Goodsell:2009xc}.

The term \eqref{kinmix} can be reabsorbed in the definition of the electric charge if the hidden photon mass is exactly zero, effectively removing hidden photons ($\gamma^\prime$s) from the theory, forbidding $\gamma\to \gamma'$ oscillations.
This is, however, not the case if there are light states charged under the local hidden U(1) symmetry associated with the
hidden photon. In this case, these particles acquire a small electric charge~\cite{Holdom:1985ag},
\begin{equation}
e Q = e_h \chi ,
\end{equation}
where $e_h$ is the unit charge of the hidden U(1), and their radiative effects in a background magnetic field still allow
$\gamma\to \gamma'$ transitions~\cite{Gies:2006ca,Ahlers:2007rd}.

In this paper we report on the results of a search for photon oscillations into general weakly coupled particles at DESY: the ``Any-Light-Particle-Search'' (ALPS).
Our experiment is based on the ``light shining through a wall'' concept~\cite{Okun:1982xi,Anselm:1986gz} 
: 
laser light is shone through a background magnetic field onto a wall.
We are searching for photons which seemingly made it through the wall by double conversion, $\gamma\to {\rm WISP}\to \gamma$,
through the WISP intermediate state.

\section{Experimental Setup}

The ALPS experiment has been already described in detail in a previous publication~\cite{Ehret:2009sq}. 
In this section we summarize the experimental setup and give details of the crucial upgrades that have made possible the physics results we present in this paper.

A simplified schematic view of the ALPS experiment is shown in Fig. \ref{fig:setup}. 
Our light source is the MOPA laser system producing up to 35 W of 1064 nm laser light~\cite{Fre07} used by us previously~\cite{Ehret:2009sq}.  
To adapt to the detector efficiency we double the frequency of the beam with a non-linear PPKTP crystal. 
This beam is directed into a vacuum pipe in which photon-WISP conversions could occur. 
Inside the pipe, an optical resonator is used to increase the laser power, enhancing proportionally a hypothetically WISP flux.
Any WISPs produced will traverse a thick light absorber (henceforth, the ``wall'') and enter into a second vacuum pipe where they can reconvert into photons.
The WISP production and photon regeneration pipes are inserted from both sides of a HERA superconducting dipole magnet.
The HERA dipole provides a magnetic field of $5$~T in a length of $8.8$~m (yellow in Fig.~\ref{fig:setup}), and its presence allows photon conversions into WISPs with spin different from one, such as axion-like particles.
Photons reconverted from WISPs appear in the regeneration tube (dashed line in Fig.~\ref{fig:setup}) with the same beam characteristics as the photons they are originating from, i.e. with the same frequency and the same TEM$_{00}$ mode to which the resonator is locked.
The end of the regeneration tube is connected to a light-tight box in which signal photons are redirected by a mirror into a lens (focal length $f=40\,$mm) which focuses the light onto a $\approx 30~\mu$m diameter beam spot on our CCD camera.

The crucial upgrades with respect to the previous setup \cite{Ehret:2009sq} are: the incorporation of a \emph{resonant} second harmonic generation scheme to increase the available \unit[532]{nm} laser power, a higher power build-up in the production resonator, the system for tuning the photon index of refraction in the production/regeneration pipes (which affects photon$\to$WISP oscillations, see below) and the incorporation of a new CCD with reduced noise and a higher quantum efficiency.

\subsection{Resonant SHG}
In order to increase the available \unit[532]{nm} laser power, a folded ring shaped resonator was built around the nonlinear PPKTP crystal (periodically poled KTiOPO$_4$) used for the second harmonic generation\footnote{The basics of second harmonic generation are extensively discussed in~\cite{Boyd:1968}. } (SHG). 
With the help of the photodiode PD1 in Fig.\ref{fig:setup} and an electronic feed-back loop the resonator length is changed in order to keep it resonant with the incident infrared laser light. 
Simultaneously the production cavity is kept in resonance by PD2 and another feedback loop acting on the frequency of the primary laser (see next section).

In its working point for long-term operation the resonant SHG emits \unit[5]{W} at a wavelength of \unit[532]{nm} from an incident power of \unit[10]{W} at \unit[1064]{nm}. At the input of the production resonator \unit[4.6]{W} were constantly available with a very good time stability of typically better than \unit[2]{\%}.

High intensities of the fundamental and/or harmonic frequencies inside the crystal cause problematic effects like thermal dephasing, nonlinear absorption processes and gray-tracking~\cite{Wang:2004,Louchev:2005}. 
The advantage of a resonant SHG is that the same amount of converted power can be achieved with a larger fundamental waist inside the crystal. The reason for this is the nonlinear nature of the SHG process. It causes the power build-up to have a linear effect on the circulating fundamental power while showing a quadratic effect on the second harmonic power.

The optimum waist size\footnote{Obtained by maximizing the Boyd-Kleinman integral.} for a non-resonant SHG with the \unit[2]{cm} long PPKTP crystal used in the ALPS setup is \unit[25]{$\mu$m}. 
For our resonant SHG, we chose a  \unit[220]{$\mu$m} fundamental waist, 9 times bigger than the single-pass optimum. A numerical simulation of this resonator results in intensity reduction factors of 15 for the fundamental and 100 for the second harmonic beam\footnote{Since its waist is always a factor of $\sqrt{2}$ smaller than the fundamental, its intensity decreases quadratically with the fundamental waist size.}.
This holds true as long as the cavity is impedance matched and the parasitic roundtrip losses are significantly smaller than the roundtrip losses due to frequency conversion.
A more detailed explanation of this simulation can be found in~\cite{Meier:2010ah}.

In this way we reduced the above mentioned problematic effects and still achieved a long-term constant output power of \unit[5]{W} behind the SHG
(monitored by PD3, Fig.\ref{fig:setup}). Green output powers above \unit[8]{W} again caused gray-tracking in our setup. Further enlargement of the fundamental beam waist was not possible because PPKTP crystals with heights bigger than \unit[1]{mm} were not available.

\subsection{Production resonator}
In our previous setup, the mirrors of the optical cavity were situated outside the production side vacuum tube, so that the green laser beam had to traverse two glass windows two times in one resonator round trip. 
Absorption and scattering in these windows (although AR-coated) limited the achievable power build-up. 
By placing the mirrors inside the vacuum, the internal losses of the production resonator in the current setup were reduced by roughly an order of magnitude, boosting the power build-up by the same factor.
The green light generated in the SHG stage is directed through the entrance window of the cylindrical vacuum chamber onto the coupling mirror (CM, radius of curvature $R = 15$ m, $1$" diameter) of the cavity. 
Using two UHV Picomotors~\cite{tm1} the CM mount is adjustable from the outside. 
Due to degradation during cavity operation (see below), two CM mirrors with transmission coefficients of $T = 1.6\times 10^{-3}$ resp. $T = 1.4\times 10^{-3}$ and two cavity end mirrors (EM, $R=\infty$) with $T = 8.8\times 10^{-5}$ resp. $T = 9.8\times 10^{-5}$ had to be used successively. 
For EM a special mirror holder was designed which is non-magnetic and suitable for high vacuum.  
With two adapted Squiggle~\cite{tm2} motors the mirror mount in the holder can be tilted remotely around two axes perpendicular to the beam and to each other. 
By guiding it in with a segmented aluminum rod, the mirror holder is slid into the laser side tube from the outer to the inner end just before the exit window. The rod carries the Teflon-coated motor wire ribbon.

These efforts resulted in a power build-up factor of \unit[$\approx 300$]{} and a continuously circulating power inside the production region of up to \unit[1.2]{kW}. This corresponds to a more than 30-fold increase compared to the setup in~\cite{Ehret:2009sq}. The power inside the cavity is monitored at all times by two different means: by measuring the incident power reflected from the CM (with PD2, Fig.\ref{fig:setup}, as explained in~\cite{Ehret:2009sq}) and by a photovoltaic cell with a sensitive area of \unit[2]{cm}$\times$\unit[4]{cm} exposed to the stray light exiting the CM chamber through a viewing port.

When operating in vacuum, the power build-up degrades after several \unit[10]{hours} of running. We have found that this is due to a slight mirror quality degradation, despite operating several orders of magnitude below the damage threshold specified by the manufacturer.

\subsection{Tuning of the refractive index}

In order to tune the index of refraction, both vacuum chambers may be closed off from their turbopumps and filled with $99.9995\%$ pure Argon. 
The pressures of both chambers are monitored with two Baratron~\cite{tm3} manometers. The pressure increases due to outgassing turned out to be smaller than 5\% in each measurement run. 
Since the refractive indexes for outgassing contaminants like water, H$_2$ or air do not surpass that of Argon, the relative change of the refractive index in the chambers due to outgassing is estimated to not surpass the relative pressure change.       

\begin{figure}[t] 
\centering
\includegraphics[width=4.2cm]{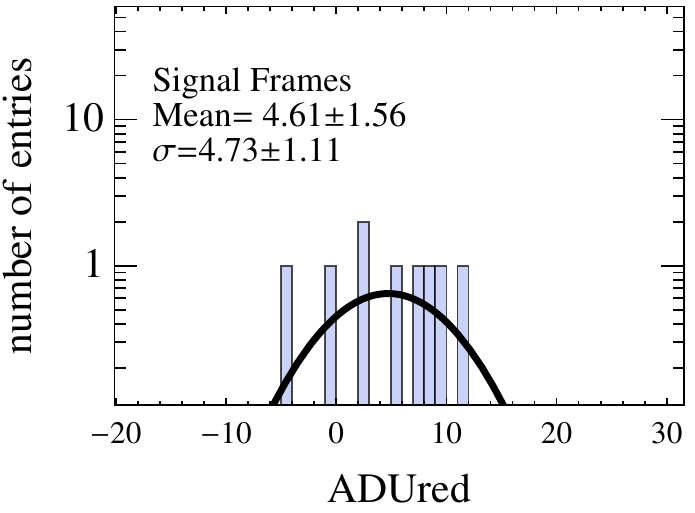} \hspace{-0.3cm}
\includegraphics[width=4.27cm]{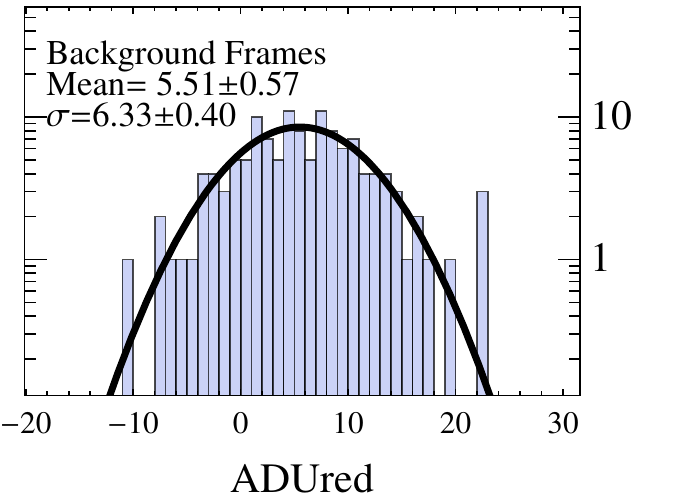}
\caption{An example of the distribution of baseline ADUred values in the signal and background sets.
The lines show Gaussians fitted to the data. 
}
\vspace{-0.2cm}
\label{fig:SB}
\end{figure}

\subsection{CCD camera and detector bench}
As detector we use a commercial CCD camera PIXIS 1024B~\cite{PIXIS} with an 1024$\times$1024 array of $(13\mu\rm m)^2$ sized pixels. 
It operates at $-70$  $^o$C and has 96\% quantum efficiency at 532 nm.
The dark current and read-out are 0.001 $e^{-}$/pixel/s and 3.8 $e^{-}$/pixel RMS, respectively.
The diameter of the focused beam is $\sim 30 \mu$m which would cover several pixels.
In order to lower the readout noise, we binned groups of $3\times 3$ pixels for the readout.
By doing this, $(87\pm5)\%$ of the light arriving to the detector is contained in the same $42\times 42\mu{\rm m}^2$ binned pixel.
Henceforth we will refer as pixels to what are actually bins of pixels.
For exposures longer than $\approx \frac{1}{2}$ hour the dark current noise dominates the total noise of the CCD. Hence, to not spoil the sensitivity of the camera, the exposure times have to be well above 30 minutes. We took 1 hour frames, because with much longer exposures the probability of spurious signals from cosmics or radioactivity close to the expected beam spot region rises. If such signals are observed the data frame is omitted.

Special care was devoted to a stable and rigid construction of the detector breadboard containing the light-tight box and the camera, for it has to be removed for accessing the wall inside the magnet. After mounting or dismounting the wall, the detector bench has to be repositioned very precisely to maintain the alignment on the pre-selected pixel. 
With numerous tests the error in repositioning of the beam spot position on the CCD was determined to be smaller than $6 \mu$m.
A further cross check of the alignment stability is provided by a reference beam which, being extracted from the back of the mirror that guides the laser into the cavity, is conducted through a pipe parallel to the HERA magnet to be finally redirected and focused onto another pixel of the CCD.
Displacements of the reference beam spot in the CCD are used to signal maladjustments in those elements, which would also alter the expected position of the spot created by LSW.

\section{Methodology}

\begin{figure}[tbp] 
\centering
\includegraphics[angle=90,width=8cm]{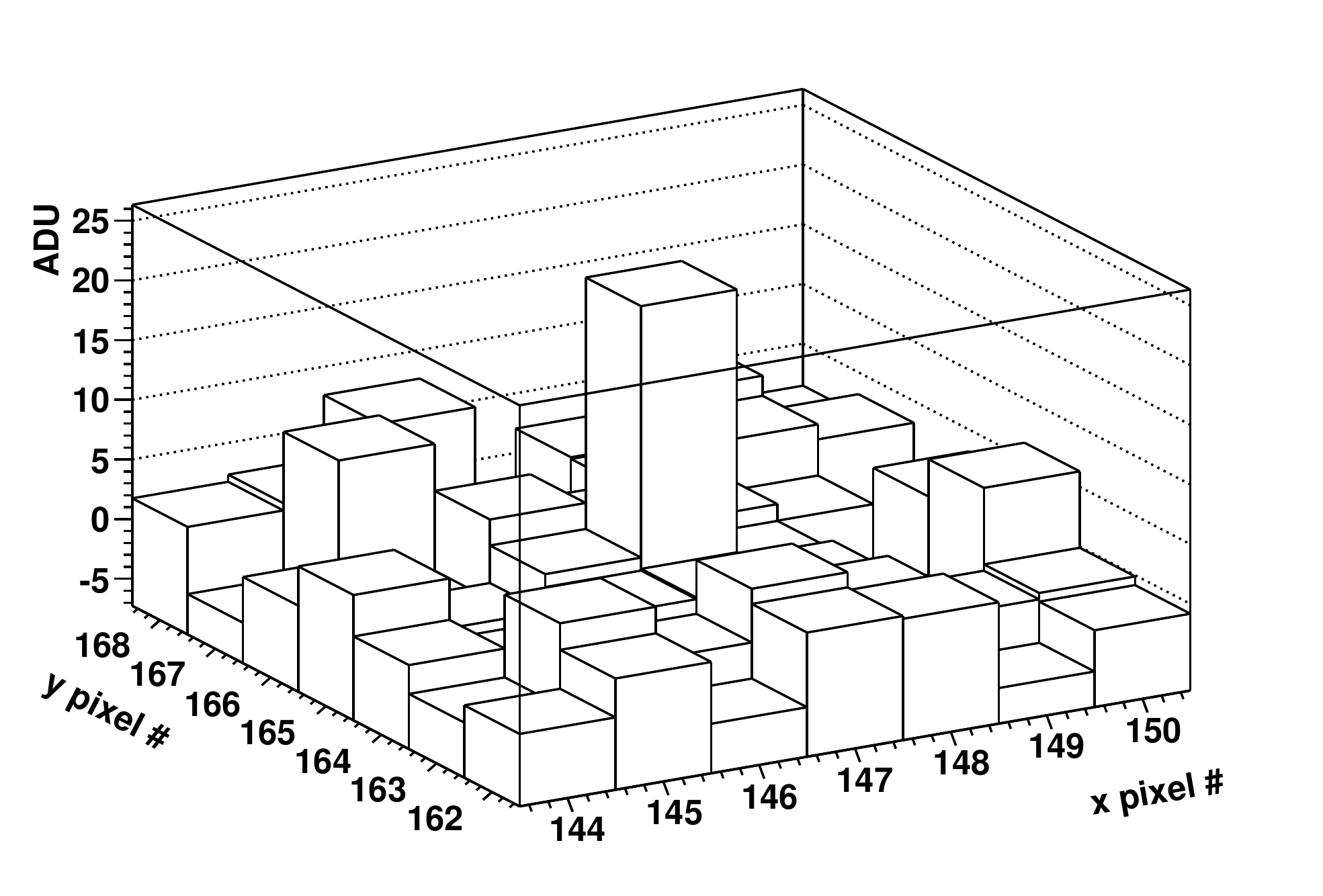}
\caption{The difference of the mean ADUred values in signal and background frames in the low-intensity test of our experiment. The signal pixel excess of $23\pm 3$ ADUs is evident in the central bin.}
\label{fig:lowintensity}
\end{figure}

The any-light-particle search proceeds the following way. After aligning the beam onto a desired pixel and repositioning the blocking wall we take 1-hour exposures (frames).
Immediately after each frame we open the reference beam shutter to test the reference beam pointing stability.
After the last frame of the session we remove the wall and check that the position of the signal pixel has not moved.
To characterize the background of the CCD we take a large number of frames with the laser turned off so that the statistical accuracy is always limited just by the number of the signal frames.

The footprint of LSW is searched for by comparing the ADU values of the signal pixel in the set of signal frames with those of the background set (1 ADU$\equiv 1e^-$).
Before the comparison, however, a selection of frames and a correction for frame-dependent baseline fluctuations is performed.
Every frame is checked for hot pixels and spurious signs from cosmics or radioactivity in a region of 25$\times$25 pixels around the signal pixel.
Frames with such activities are omitted from the data analysis.
The fluctuations are accounted for by subtracting the average of a $11\times 11$ array of pixels around the signal pixel from the ADU value of the signal pixel. 
This difference is called ADUred (for ADU reduced) in the following. 

The LSW flux can be estimated by comparing the mean of the distribution of ADUred values in the sets of signal and background frames. 
For every distribution, we check that the width corresponds to the value expected from dark current and read-out noise.
An example of the resulting distributions is shown in Fig.~\ref{fig:SB}.
Finally, in the absence of any evidence for LSW, a confidence level is obtained by the method of Feldman and Cousins~\cite{Feld98}.

As a direct test of the whole setup and methodology we took a set of 4 frames \emph{without the wall} and a set of attenuators placed in the way of the beam to the CCD. The expected ADUred signal level per frame was calculated to be within $5$ and $50$~mHz.
The distribution of the difference of the mean ADUred values in signal and background frames is shown in Fig.~\ref{fig:lowintensity}.
The signal pixel shows an evident excess of $23.3\pm 3.2$~ADUs which translates into a flux of ($7.9\pm 1.2$)~mHz
when we take into account our full collection efficiency of $(82\pm 5)\%$. This result is in excellent agreement with expectations and proves the high sensitivity of the apparatus.

\section{Results and Interpretation}

\begin{table*}[tdp] \footnotesize
\caption{Summary table of our data sets and results.
Data sets were taken with magnet on or off, laser polarization parallel or perpendicular to the magnetic field, and different Argon pressures in the production/regeneration tubes, corresponding to different refractive indexes, $n$. The number of frames is separated for different signal positions on the CCD. Also shown is the average laser power in each collection of sets and the $95\%$ limits on the conversion probability. Finally we show the different WISPs probed for in each configuration.
The 27 frames mentioned in row 5 are the same 27 frames collected with magnet on and without Argon.}
\begin{center}
\begin{tabular}{|c|c|c|c|c|c||c|c|c|c|c|}	\hline														
Magnet	& Laser Pol.	& \# frames	& Pressure/mbar & $n-1$& Power/W & Prob. 95$\%$ limit 	& $0^-$	& $0^+$	& $\gamma'$	& MCP  \\
\hline
On	& Par.	& 5/6 &	<$10^{-5}$	& 0 & 1096& 	2.25$\times10^{-25}$	& \checkmark	& -	& -	&\checkmark		\\
On	& Par.	& 8	&	0.18	& $5.0\times10^{-8}$ & 1044	&  10.8$\times10^{-25}$	& \checkmark	& -	& -	&\checkmark			\\
\hline
On	& Perp.	&9/5/2&	<$10^{-5}$ & 0 &1088& 2.08$\times10^{-25}$&	- & \checkmark & -	&\checkmark	\\			
On	& Perp.	& 8	&	0.18 & $5.0\times10^{-8}$ & 954	 & 5.22$\times10^{-25}$	& -	& \checkmark			& -	&\checkmark	\\			
\hline
Off (1+3)	& Par.+Perp.		&27	&	<$10^{-5}$	& 0 & 1121	&  1.14$\times10^{-25}$ & -	& -			&\checkmark	&-						\\
Off	& Par. & 9	&0.11/0.14 & $3.1/3.9\times10^{-8}$ & 968	& 	3.11$\times10^{-25}$	& -	& -			&\checkmark	&-				\\
\hline
\end{tabular}
\end{center}
\label{Tab:results}
\end{table*}%

The LSW probability $P(\gamma\to{\rm WISP}\to\gamma)$ factorizes into the product of the probabilities for photon $\rightarrow$ WISP conversion before the wall and for WISP $\rightarrow$ photon conversion behind it. The latter are given by the same expression if CPT is conserved in the transitions -- an assumption that holds in the cases considered in the following -- 
but depend upon the WISP under consideration.
In order to cover all the interesting candidates, we took frames in three different configurations: with polarization parallel to the magnetic field, perpendicular to it, or with the magnetic field turned off. Moreover, in each of these three configurations data has been collected both 
with vacuum and with a small amount of Argon gas in the production/regeneration tubes, respectively, to realize refractive indices different from unity.
In Table~\ref{Tab:results} we show a summary of all our data sets.

\begin{figure*}[ht]
\vspace{-0.6cm}
\centering
\includegraphics[width=9cm]{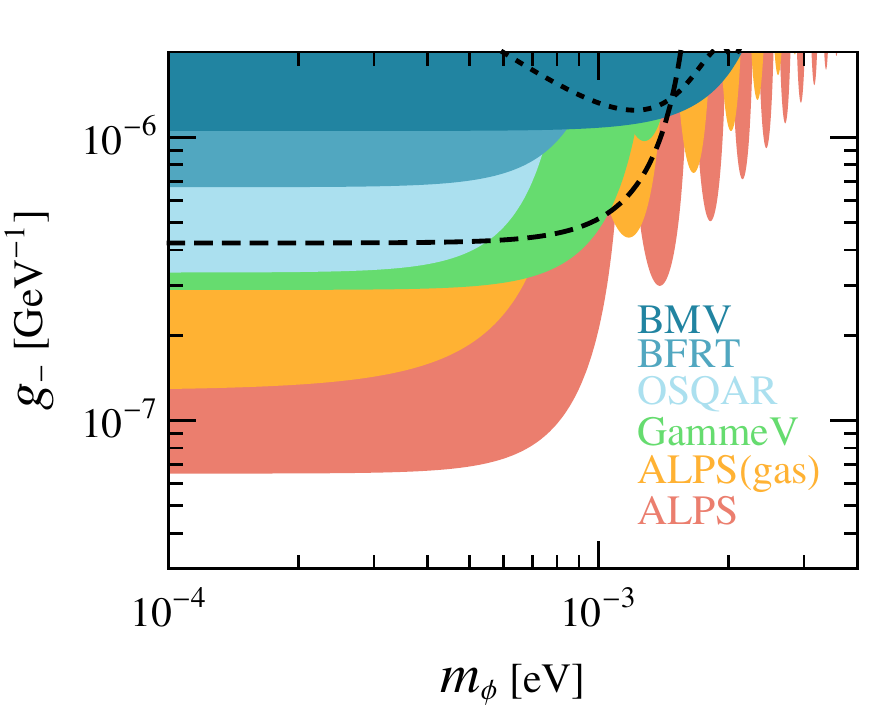} \includegraphics[width=9cm]{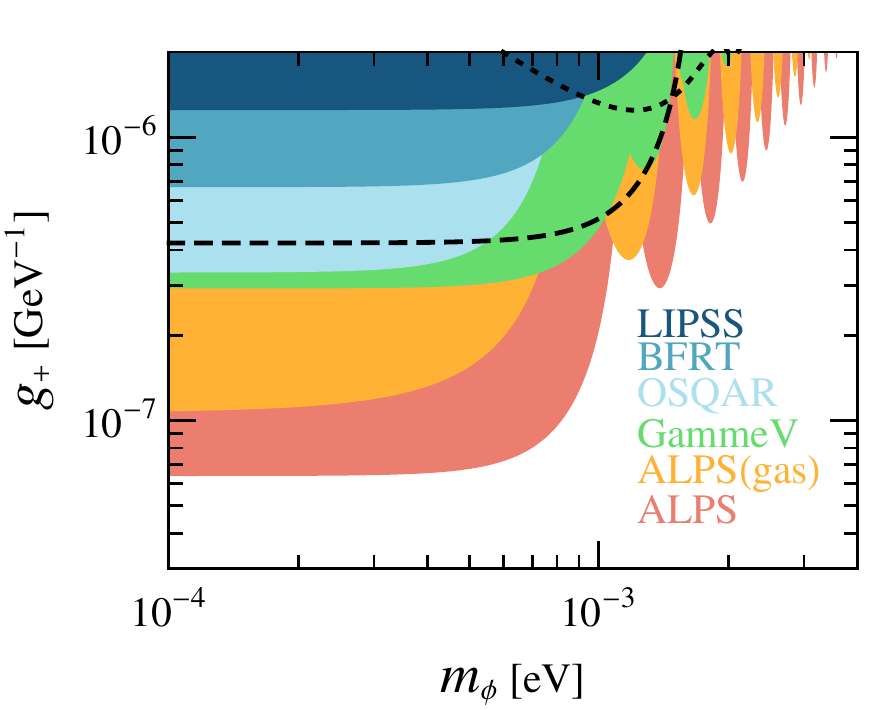}
\vspace{-0.2cm}
\caption{Exclusion limits ($95\%$ C.L.) for pseudoscalar (left) and scalar (right) axion-like-particles as described in this paper from the
vacuum and gas runs.
Shown for comparison are results from the BMV~\cite{Fouche:2008jk,Robilliard:2007bq}, BFRT~\cite{Cameron:1993mr}, GammeV~\cite{Chou:2007zzc}, LIPSS~\cite{Afanasev:2008jt} and OSQAR~\cite{Pugnat:2007nu} LSW experiments. 
Dashed and dotted lines show the bounds on ALP-induced dichroism and birefringence from the PVLAS experiment~\cite{Zavattini:2007ee}.}
\label{fig:alps}
\end{figure*}

ALPs couple to two photons, cf. Eqs.~\eqref{ALPcoupling_pseudoscalar} and \eqref{ALPcoupling_scalar}.
Correspondingly, $\gamma\to$ALP conversions may occur in our case only if the magnetic field is switched on.
The conversion probability is then given by
\begin{eqnarray}
\label{axionprob}
P(\gamma \leftrightarrow \phi^{(-)}  )= 4\frac{(g_{_-} \omega  B \cos\theta)^2}{M^{4}}\sin^2\left(\frac{M^{2} l }{4\omega}\right),
\end{eqnarray}
where $\omega$ is the photon energy, $B$ and  $l$ are the magnitude and the length, respectively, of the magnetic field
region, $\theta$ is the angle between the laser polarization and the direction of the magnetic field (assumed to be transverse to the direction of the photon's motion), and $M^2$ is the difference of photon and WISP effective squared masses,
\begin{equation}
\label{eff_mass}
M^2= m_\phi^2+2\omega^2(n-1) ,
\end{equation}
with $n$ the photon refractive index.
For a scalar ALP $\phi^{(+)}$, $g_- \cos\theta$ has to be replaced by $g_+ \sin\theta$
in \eqref{axionprob}.
With the help of Eq.~\eqref{axionprob} our limits on the probabilities
of light shining through a wall from Table~\ref{Tab:results} are translated into limits of the couplings
$g_\pm$ vs. ALP mass $m_\phi$, cf. Fig.~\ref{fig:alps}.

The most stringent constraints can be obtained for vacuum conditions in the beam pipes ($n\equiv 1$) since
for $n>1$ the photon effective mass $-2\omega^2(n-1)$ suppresses the conversions.
For this reason we concentrated most of our measurement time in this case.
For larger masses, the $\gamma\to$ALP conversions increasingly lose coherence and there are even regions in coupling
vs. mass which are unconstrained by our vacuum measurements.
These regions correspond specially to masses where $m_\phi^2 l/(4 \omega)=\pi\times s$, where $s$ is a \emph{nonzero} integer.
These gaps in sensitivity can be filled by introducing an adequate amount of gas in the WISP production and photon regeneration regions. 
The optimal index of refraction can be computed by maximizing Eq.~\ref{axionprob} with respect to $n-1$.
The maxima are the infinite but numerable solutions of the transcendental equation
\begin{eqnarray}
\tan y &=& \frac{y}{2} \quad {\rm with}\quad y=\frac{M^2L}{4\omega}  \ , \\
 \left\{y\right\} &=& \left\{y_0,y_1...\right\}=\left\{0,4.27,7.60,...\right\} .
\end{eqnarray} 
For a mass corresponding to a gap, specified by the integer $s$, we have  $y_{s}=M^2L/(4\omega)=s\pi + \omega L(n-1)/2$. 
The optimum index of refraction is given by $n-1= 2 (y_{s}-s\pi)/(\omega L)$. 
In our setup, $l=4.3$ m and $\omega=2.33$~eV, so the lowest mass gap in our experiment, $s=1$, can be optimally covered by using 
$(n-1)\simeq 4.45 \times 10^{-8}$.
As $s$ grows, $y_s\to s \pi +\pi/2$ and the required index of refraction to cover all the high mass gaps becomes independent of $s$,  $(n-1)\simeq 6.2 \times 10^{-8}$.
We determined numerically that an intermediate value of $(n-1)\simeq 5 \times 10^{-8}$ is a good compromise to cover all the gaps shown in Fig.~\ref{fig:alps}. This was achieved by introducing Ar gas at a pressure of $0.18$~mbar.
We must emphasize that this is not equivalent to restoring $\gamma$-ALP coherence, which would 
correspond to make $M^2\to 0$, but to increase the $\gamma$-ALP relative phase velocity to have an extra half oscillation length (an extra phase of $\sim \pi/2$) in our available magnetic field length.

Our limits on ALPs, cf. Fig.~\ref{fig:alps}, are the most stringent laboratory bounds in the sub-eV mass range.
Stellar evolution considerations~\cite{Raffelt:1996wa} provide often the most prominent constraints on WISPs. 
For instance, stellar population surveys in globular clusters or searches for solar ALPs~\cite{Andriamonje:2007ew} can constrain the ALP parameter space up to two orders of magnitude further. 
However, as pointed out in many papers~\cite{Jaeckel:2006xm} such constraints can be evaded in certain WISP models, an observation that renders our constraints relevant. 
Hidden photons, to which we shall come next, are one of these models in which a WISP can easily evade the astrophysical bounds.

\begin{figure}[tt]
\vspace{0cm}
\centering
\includegraphics[width=9cm]{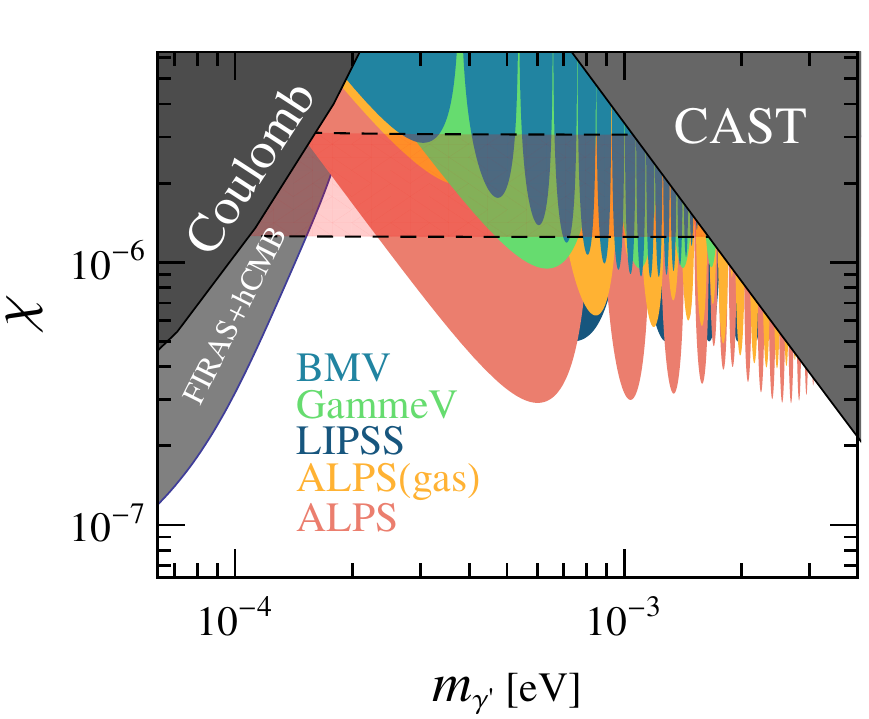}
\vspace{0cm}
\caption{$95\%$ C.L. exclusion limits for hidden photons.
For comparison we show the exclusion limits from the BMV~\cite{Fouche:2008jk}, GammeV~\cite{Ahlers:2007qf} and LIPSS~\cite{Afanasev:2008jt} experiments.
Also shown are limits from searches of modifications of Coulomb's law~\cite{Bartlett:1988yy}, distortions of the CMB spectrum~\cite{Jaeckel:2008fi} and the solar axion search by CAST~\cite{Redondo:2008aa}.
Hidden photons in the horizontal redish band could account for the apparent excess in the relic neutrino density recently reported by WMAP-7~\cite{Komatsu:2010fb}.}
\label{fig:hp}
\end{figure}

The oscillations into  hidden photons occur also in the absence of a magnetic field.
The probability is given by~\cite{Ahlers:2007qf}
\begin{equation}
\label{hpprob2}
P(\gamma \leftrightarrow \gamma^\prime  ) \simeq
4\chi^2 \frac{m_{\gamma^\prime}^4}{M^4}\sin^2\left(\frac{M^2 l }{4\omega}\right) ,
\end{equation}
where $L$ is the propagation length and $M^2$ is the same as in Eq.~\eqref{eff_mass}, with the replacement
$m_\phi\rightarrow m_{\gamma^\prime}$.
Clearly, the conversion probability vanishes for $m_{\gamma^\prime}\to 0$. This is
also apparent in Fig.~\ref{fig:hp} which displays our limit on the kinetic mixing $\chi$ vs. $m_{\gamma^\prime}$. 
Note that the lengths available for the production and regeneration processes are different in this case. 
They are given by the sizes of vacuum regions and not by the magnetic field, and they are different from each other, cf. Fig.~\ref{fig:setup}. 
As a consequence, the optimal values of the refraction index (calculated along the lines described for ALPs) that cover the gaps are also different from each other, cf. Tab~\ref{Tab:results}.

ALPS clearly sets the most severe constraints in the $\sim $~meV mass
region, not only as compared to other laboratory experiments such as searches for deviations from 
Coulomb's law~\cite{Bartlett:1988yy}, but also to inferences exploiting cosmology or astrophysics, 
notably solar energy loss considerations and searches for
solar hidden photons~\cite{Redondo:2008aa}. 
Interestingly, a hidden photon in this mass region could appear
as an extra contribution to the cosmic radiation density during the epoch after
big bang nucleosynthesis and before recombination~\cite{Jaeckel:2008fi} -- a puzzling possibility supported (albeit with less than $2\sigma$ significance) by the recent analysis of WMAP data~\cite{Komatsu:2010fb} leading to a value of the effective 
number of neutrinos higher than the standard value of three by an amount $\Delta N_{\nu}^{\rm eff}=1.3\pm 0.9$.
As seen from Fig.~\ref{fig:hp}, the new data from ALPS exclude this possibility nearly entirely, up to 
a tiny region in parameter space, $m_{\gamma^\prime}\approx 0.18$~meV, $\chi\approx 1.4\times 10^{-6}$.
An easy way to probe for hidden photons within this region\footnote{Such hidden photons intriguingly would allow for long distance 
communications through oceans or even the Earth~\cite{Jaeckel:2009wm}} 
is to prolong the lengths of the production/regeneration
tubes in a future experiment.

\begin{figure}[ht]
\vspace{0cm}
\centering
\includegraphics[width=9cm]{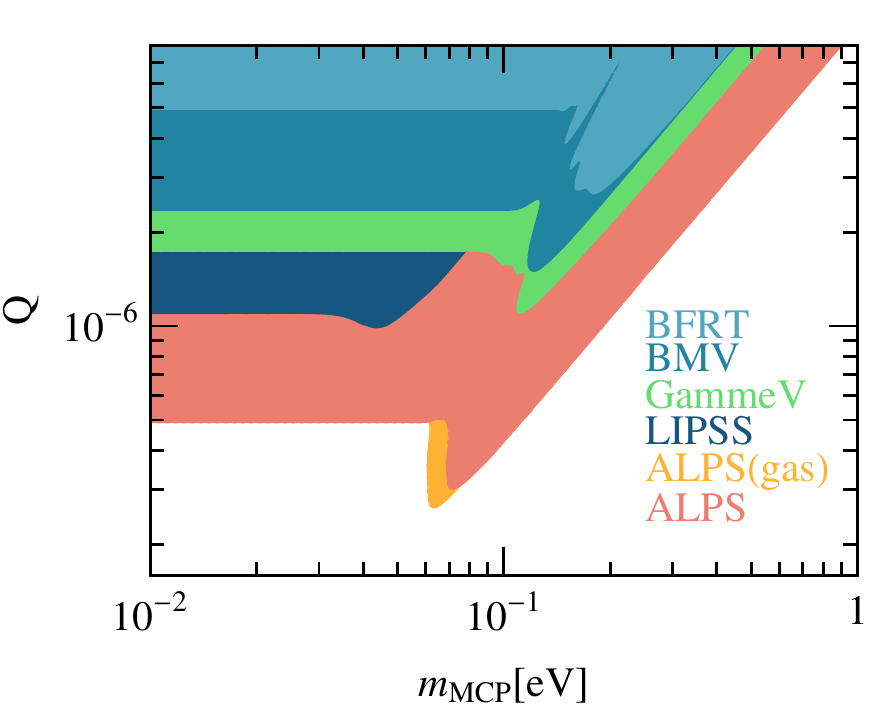}
\vspace{0cm}
\caption{$95\%$ C.L. exclusion limits for mini-charged particles in models with a hidden photon. We have taken 
$e_h=e$, so $Q=\chi$.}
\label{fig:mcps}
\end{figure}

As mentioned in the introduction, even in the $m_{\gamma^\prime}=0$ case, $\gamma\to \gamma'$ oscillations are possible in a magnetic field if there are light particles charged under the hidden U(1), i.e. mini-charged particles (MCPs).
The corresponding conversion probability is~\cite{Burrage:2009yz}
\begin{equation}
\label{hpprob1}
P(\gamma \leftrightarrow \gamma^\prime  )\simeq
\left| \frac{m_\phi^2}{M^2}\right| \left|e^{i k_+ L}-e^{i k_- L}\right|^2 , 
\end{equation}
\begin{equation}
k_\pm = \frac{1}{4\omega}\left(2\omega^2 (n-1)-m_\phi^2 \pm M^2 \right)(1\pm \chi^2 m_\phi^2/M^2) , 
\end{equation}
where now $m_\phi$ is the hidden photon effective mass which 
can be written as $m_\phi^2=-2\omega^2 \Delta N (Q,B,m_{_{\rm MCP}})$ with \break $\Delta N (Q,B,m_{_{\rm MCP}})$ the complex index of refraction due to mini-charged particles of charge $Q$ and mass $m_{_{\rm MCP}}$.
The explicit expression for $\Delta N$  for scalar or Dirac spinor MCPs is given in Ref.~\cite{Ahlers:2007rd}. 
Our limits on the charge of mini-charged particles vs. their mass are displayed in Fig.~\ref{fig:mcps}, where we have assumed a Dirac MCP and $e_h=e$ for simplicity\footnote{For smaller values of $e_h$, like the ones arising in hyperweak scenarios~\cite{Goodsell:2009xc}, our bounds get worse, see~\cite{Ahlers:2007qf}.}. 
Since there are no gaps in our sensitivity we made no dedicated gas runs. However, the gas data collected for ALPs can be also analyzed in terms of MCPs. In this case there is a resonance region where our gas bounds improve the vacuum bounds. 
This happens in the region where the real part of the effective mass squared of the hidden photons is negative and can match that of the laser photons. 
However, the resonance is cut-off by the imaginary part so that finally the improvement is only modest.

Also in this case the new constraints from ALPS give the best laboratory bounds on 
$\sim$eV mass mini-charged particles. At larger masses our bounds decrease very fast and are soon covered by the constraints obtained from Lamb shift 
measurements~\cite{Gluck:2007ia} and from searches for an invisible decay of 
orthopositronium~\cite{Badertscher:2006fm} (both at the $Q \lesssim 10^{-4}$ level).
At extremely small masses ($\lesssim 10^{-6}$eV), searches for deviations from Coulomb's law surpass our limit slightly~\cite{Jaeckel:2009dh}.
However, like in the ALPs case, also here the bounds
from astrophysics and cosmology are much stronger: the lifetime of red giants, e.g., exclude a 
charge larger than $Q\gtrsim {\rm few}\times 10^{-14}$, for $m_{\rm MCP}\lesssim $~MeV~\cite{Davidson:2000hf}.

\section{Conclusions}

In this experiment we have set limits on the existence of Weakly Interacting Sub-eV Particles (WISPs) such as Axion-Like Particles (ALPs), hidden photons and mini-charged particles with a light-shining-through-a-wall setup. 
These are the most stringent purely laboratory constraints up to date. 
Furthermore, we can almost completely rule out the possibility of explaining the current trend of WMAP and large-scale-structure probes of a non-standard radiation density contribution due to hidden photons. 

Our setup uses extensively optical resonators to maximize the photon flux available for WISP searches. 
It is increasingly evident that this is the way to go for a new generation of laboratory experiments looking for WISPs.
In order to compete with the strong astrophysical considerations that normally challenge the existence of WISPs the next generation will have to push four basic frontiers: stronger magnets, more laser power in the production region, more sophisticated low-background detectors and the implementation of the promising resonant regeneration technique~\cite{Hoogeveen:1990vq} worked out in experimental details recently~\cite{Mueller:2009wt,Meier:2010ah}. 

These efforts could not await a more exciting reward. 
For instance, testing hidden photons with $\chi\lesssim 10^{-8}$ for $10^{-6}\,{\rm eV} \lesssim m_{\gamma^\prime}\lesssim 10^{-2}\,{\rm eV}$ would mean entering in a territory where hidden photons could signal properties of the extra dimensions predicted by string theory~\cite{Goodsell:2009xc}. 

Furthermore, reaching sensitivities to the ALP photon couplings in the range $g_\pm \lesssim 10^{-11}$~GeV$^{-1}$ for $m_\phi \lesssim 10^{-9}$~eV would allow to definitively test the ALP interpretation of the puzzling astronomical observations mentioned in the introduction. Moreover, further experimental efforts in this direction could end up showing a clear future road-map to even test QCD axions with $\sim$meV masses.

\section*{Acknowledgments}
The ALPS collaboration thanks the MVS and MKS groups at DESY for crucial support, especially
H.~Br\"uck, J.~Fischer, E.~Gadwinkel, C.~Hagedorn, H.~Herzog, K.~Pr\"uss and M.~Stolper.
We acknowledge financial support from the Helmholtz Association and from the Centre for Quantum Engineering and Space-Time Research (QUEST).
JR acknowledges support from the SFB 676 and the DFG cluster of excellence EXC 153 Origin and Structure of the Universe.

\bibliographystyle{utcaps}


\providecommand{\href}[2]{#2}\begingroup\raggedright\endgroup

\end{document}